# EXPLAINABLE AND SCALABLE MACHINE-LEARNING ALGORITHMS FOR DETECTION OF AUTISM SPECTRUM DISORDER USING FMRI DATA

**Taban Eslami, Joseph S. Raiker, Fahad Saeed**
Correspondence Email: fsaeed@fiu.edu
School of Computing and Information Sciences
Florida International University (FIU), Miami FL

**Abstract**

Diagnosing Autism Spectrum Disorder (ASD) is a challenging problem, and is based purely on behavioral descriptions of symptomology (DSM-5/ICD-10), and requires informants to observe children with disorder across different settings (e.g. home, school). Numerous limitations (e.g., informant discrepancies, lack of adherence to assessment guidelines, informant biases) to current diagnostic practices have the potential to result in over-, under-, or misdiagnosis of the disorder. Advances in neuroimaging technologies are providing a critical step towards a more objective assessment of the disorder. Prior research provides strong evidence that structural and functional magnetic resonance imaging (MRI) data collected from individuals with ASD exhibit distinguishing characteristics that differ in local and global spatial, and temporal neural-patterns of the brain. Our proposed deep-learning model *ASD-DiagNet* exhibits consistently high accuracy for classification of ASD brain scans from neurotypical scans. We have for the first time integrated traditional machine-learning and deep-learning techniques that allows us to isolate ASD biomarkers from MRI data sets. Our method, called *Auto-ASD-Network,* uses a combination of deep-learning and Support Vector Machines (SVM) to classify ASD scans from neurotypical scans. Such *interpretable* models would help explain the decisions made by deep-learning techniques leading to knowledge discovery for neuroscientists, and transparent analysis for clinicians.

## 1 Introduction

Current nosological approaches to characterizing mental health problems rely on largely overlapping emotional, behavioral, and cognitive features creating considerable challenges for both researchers and clinicians in assessing and treating psychiatric disorders. Specifically, misdiagnosis as well as the provision of treatments intended for one disorder to those impacted by another disorder represent critical challenges to the field of mental health and highlight an urgent need for the identification of appropriate biomarkers as well as novel approaches to aid clinicians in diagnosing mental illness and monitoring treatment outcomes [1]. The current psychiatric diagnostic process is based nearly exclusively on reports of behavioral and/or emotional observations of symptomology (DSM-5/ICD-10) and requires careful consideration of multiple aspects related to descriptions of these observations by informants such as parents, teachers, or children (e.g., when symptoms started, what they look like over time, how long they have lasted). Collectively, all of these factors rely heavily upon: a) clinicians obtaining sufficient information to characterize the nature of symptomatology and b) patient's and/or other informant's (e.g., parents, teachers) ability to sufficiently describe these symptoms as well as recall their onset, course, and duration. Unfortunately, in contrast to other health conditions (e.g., diabetes,

HIV, hepatitis-C) which can be characterized based on interpretation of quantitative tests of biological markers (e.g., blood tests), no such quantitative tests of biological processes currently exist for mental health conditions such as ASD [2]. Fortunately, the emergence of "big data" analytics and the increasing availability of data shared across various repositories – particularly in the case of federally funded research - provide a renewed opportunity for leveraging more advanced analytic approaches (e.g., machine learning, deep learning) to develop diagnostic aids for clinicians as well as isolate potential biomarkers for a number of these mental health conditions (e.g., ADHD, ASD).

Autism Spectrum Disorder (ASD) represents a lifelong neuro-developmental disorder characterized by deficits in social communication, social interaction, and stereotypical interests and/or repetitive behaviors that emerge in the early developmental period [3]. Recent estimates indicate that approximately 62 in every 10,000 (1/160) children worldwide are impacted by the disorder [4] with the disorder being more common in males relative to females [5]. In the most recent DSM, ASD is conceptualized as reflecting a broad spectrum of symptoms that present to varying degrees across the domains of social communication and interaction as well as stereotypical interests and/or repetitive behaviors [3]. Notably, children meeting formal diagnostic criteria for ASD under the current diagnostic system may have been characterized as having one of several potential diagnoses (e.g., Asperger's Disorder, Autism, Childhood Disintegrative Disorder) prior to 2013 with the publication of the Diagnostic and Statistical Manual – $5_{th}$ Edition (DSM-5). In addition to the reclassification of these formerly distinct disorders into a broader umbrella diagnostic category termed Autism Spectrum Disorder (ASD), the ability to confer diagnostic comorbidities with other disorders was also expanded in the DSM-5. For example, prior to DSM-5, clinicians were unable to confer a diagnosis of ADHD in a child also diagnosed with Autism or Asperger's Disorder; however, this exclusion no longer applies resulting in substantial work on shared underlying vulnerabilities between the two disorders [6]. Collectively, these changes reflect our evolving understanding of the shared and distinct behavioral features that characterize ASD and highlight the need for further inquiry into understanding their neurobiological underpinnings and identifying more optimal approaches for assessing and diagnosing ASD.

## Current diagnostic practices

ASD can be diagnosed as early as 2 years of age and earlier detection is considered critical to improving overall long-term prognosis in those with the disorder [7]. Specifically, recent emphasis on the importance of early detection and intervention has resulted in substantial work to better understand optimal windows for identification and intervention particularly in very early childhood [8]. Current clinical practice guidelines recommend the administration of empirically validated screening instruments universally to all young children in an effort to identify children who may be at risk for a diagnosis of ASD (e.g., Autism Screening Questionnaire; Childhood Autism Rating Scale; [9]). Subsequently, in the presence of a score indicating elevated risk for a diagnosis of ASD, these guidelines recommend the use of a more comprehensive assessment (e.g., structured or semi-structured interviews) to assess the onset, course, duration, and response to treatment of significant ASD-related symptoms. Finally, follow-up via multidisciplinary consultation with other specialists (e.g., genetic testing, auditory evaluation, physical examinations) to rule out other comorbidities or conditions that may account for the symptoms as well as facilitate treatment planning are also recommended [9]. While a number of rating scales and interviews to evaluate ASD-related difficulties have been developed over the years, their widespread adoption by clinicians may be potentially hindered by several limitations. For example, some of these instruments are relatively costly, lengthy to administer, and may require substantial time to score and interpret. Further, some of them also require extensive training prior to their use in clinical practice. Despite this, however, it is critical for clinicians to adhere to these guidelines until more novel assessment approaches and diagnostic aids are available given the difficulty associated with differentiating a diagnosis of ASD from other clinical disorders (e.g., ADHD, OCD, language disorders, intellectual disability) due to notable shared features across these various mental health problems.

**Social Communication:** The development of language in neurotypical children occurs early in development and exponentially increases between the ages of 6 months and 3 years [10]. Despite this, however, considerable variability exists within this age range in the extent which children develop sufficient language skills. As a result, challenges in the development of language may reflect a myriad of possible causes (e.g., slower than usual language development, diagnosable language delays, intellectual disability) independent of, or in conjunction with, a diagnosis of ASD. In addition to challenges in language development in some cases of ASD, substantial challenges in the development of other critical communication skills are also present in this population (e.g., echolalia, understanding the pragmatics of language [11] [12]). Critically, the DSM-5's introduction of the new diagnostic label social (pragmatic) communication disorder which also shares several verbal and non-verbal communication problems as ASD highlights the need for careful consideration of other features of ASD in the differential diagnosis of the disorder.

**Social Interaction:** In addition to challenges related to the development and appropriate use of language skills, children with ASD also manifest deficits in various aspects of social interactions. Broadly, deficits in social skills may reflect difficulties in the areas of joint attention, difficulty making eye contact, difficulties with developing and maintaining friendships with others, as well as lacking the understanding of social and emotional reciprocity [13]. Further, children with ASD may have difficulty engaging appropriately in conversations with other children, may interrupt at inappropriate times, and may demonstrate perseveration around topics of conversation – particularly those with which they may have great interest or enthusiasm [13]. Critically, these social skills deficits are not unique to children with ASD and occur across other mental health disorders such as ADHD [14]. As a result, the mere presence of social interaction problems must be carefully considered within the context of other symptoms and features reported by informants.

**Stereotypical interests and/or repetitive behaviors:** Finally, children with ASD may also demonstrate behavioral difficulties related to flexibility in routines, insistence on sameness, repetitive behaviors (e.g., hand flapping), excessive interests or preoccupations with certain topics or ideas, as well as excessive sensitivity to sensory stimuli (e.g., sounds, touch). Notably, some of these behaviors (e.g., engaging in repetitive behaviors, insistence on sameness) are observed in other disorders such as Obsessive-Compulsive Disorder (OCD) although the form and severity of these behaviors often varies across these disorders [15]. Similarly, their preoccupation with certain topics or ideas may be misreported by informants as reflecting obsessional thinking despite the fact that there is no underlying anxiety accompanying these thoughts like is observed in disorders such as OCD.

Collectively, symptoms of ASD may look very similar to symptoms observed in other psychiatric conditions (e.g., ADHD, OCD, intellectual disability). As a result, individuals suspected of meeting formal diagnostic criteria for ASD must be carefully evaluated using empirically-established instruments that may require extensive training, may be costly to obtain and administer, and still rely upon informant report and recall as well as appropriate clinical judgment. This combined with emerging interest in early identification and intervention highlights the need to better understand the underlying neurobiological mechanisms of the disorder as well as identify novel methods of quantifying the likelihood that an individual meets formal diagnostic criteria for ASD rather than another disorder.

**Computational Techniques to Identify ASD brain scans**

Quantitative analysis of brain imaging data can provide valuable biomarkers that may result in more accurate and expedient diagnosis of ASD. Machine learning techniques using brain imaging data (e.g. Magnetic Resonance Imaging (MRI) and functional Magnetic Resonance Imaging (fMRI)) have been extensively used by researchers in hopes of aiding in the diagnosis of brain disorders like Alzheimer's, ADHD, MCI, and Autism [17] [18] [19] [20] [21] [22] [23] [24]. For this chapter, we will define and explain the imaging data sets. This will be followed by two techniques that can be used for identifying biomarkers and classifying ASD brain scans from normal developmental scans.

**Imaging Data Sets**

Functional Magnetic Resonance Imaging (fMRI) is a non-invasive techniques that uses Blood Oxygen Level Dependent (BOLD) contrast to study functional activities of the brain [25]. Using computational techniques to analyze fMRI data for discovering hidden patterns has gained significant attention [17]. During a fMRI session, a sequence of images are taken by the scanner while the subject either does some pre-determined task (for task based fMRI) or the subject rests (for resting state fMRI). The data that is collected from the fMRI consists of voxels which are smallest addressable element in the brain. These voxels can be in millions and each voxel incorporate large number of neurons inside it. Hemodynamic changes inside the brain are revealed as intensity changes of the brain voxels [26]. By keeping track of intensity of each voxel over time, a time series is extracted out of each voxel which is used for further analysis. There is much interest in developing computational techniques that can be used for identifying and diagnosing mental disorders. Therefore, researchers have come up with gold standard data sets that can be used by computational researchers as a benchmark for the proposed algorithms and models. To this end, Autism Brain Imaging Data Exchange (ABIDE) initiative has provided with two main data sets that can be used by researchers to verify their algorithms and models that may diagnose ASD.

Autism Brain Imaging Data Exchange I (ABIDE I): The Autism Brain Imaging Data Exchange I (ABIDE I) represents data from 17 international sites, sharing previously collected resting state functional magnetic resonance imaging (rs-fMRI), anatomical and phenotypic datasets.

The whole benchmark consists of 1112 data sets from 539 individuals with ASD and 573 from typical controls. The ages of these varied between 7 and 64 years with a 14.7 years median. The establishment of this data set showed the feasibility of aggregating rs-fMRI and structural MRI (sMRI) data sets that may have been collected in different sites with different instruments and different experimental conditions and showed the strengths and limitation of these data sets. In general, there is a consensus that these data sets show the utility of the whole brain is captured and the regional properties of the ASD in terms of brain connectome is available.

Autism Brain Imaging Data Exchange II (ABIDE II): The establishment of ABIDE I data set demonstrated that such resources will be feasible and their utility for aggregating imaging knowledge about ASD across sites is beneficial. However, the complexity of the connectome and the wide variability and heterogeneity underscored that even larger data sets will be needed for better characterization of Autism Spectrum. Therefore, ADIDE II data sets have aggregated over 1000 additional data sets which have greater phenotypic characterization for core ASD symptoms. The data also include temporal information with 38 individuals who MRI images are available between 1 and 4 years. This provides greater insight into ASD data and its connectome with temporal information that might be needed for efficient classification. The data set consists of 1114 data sets from 521 ASD and 593 healthy controls and the data was collected from 19 different sites. These data sets are HIPPA compliant and have been available for public release since 2016.

**ABIDE Preprocessed:** fMRI data collection from different sites is a complex operation and deals with problems of having data acquired from different experimental conditions and different instruments. Any data collected has to be pre-processed using established pipelines that eliminate variations and noise from these data sets. Since the objective of these data sets were to make them available to wide variety of researchers; ABIDE initiative has made pre-processed data available to make these data sets consistent. Therefore, it reduces time it takes computational scientists to enter the field and not deal with problems of pre-processing that generally relates to neuroscience. The Preprocessed Connectomes Project (PCP) collaborated with ABIDE to preprocess neuroimaging data from the Autism Brain Imaging Data Exchange (ABIDE). In order to be consistent across different data sets, the ABIDE data was pre-processed by five different teams using their preferred tools namely, Connectome Computation System (CCS), the Configurable Pipeline for the Analysis of Connectomes (CPAC), the Data Processing Assistant for Resting-State fMRI (DPARSF) and the NeuroImaging Analysis Kit. Four different preprocessing strategies were performed with each pipeline: all combinations of with and without filtering and with and without global signal correction. Structural pre-processing was also provided by the ABIDE initiative using 3 different pipelines namely ANTS, CIVET, and FreeSurfer. All of the data sets along with their meta-data is available to the users via the ABIDE webpage.

ABIDE I/II data sets have been extremely helpful in developing computational methods and benchmarks for classification of ASD brain scans from neurotypical scans. However, in order to design and develop machine-learning algorithms that require a lot of data; additional techniques such as data simulation and augmentation are needed. Some details about our implemented techniques along with other possible strategies that can be used by the users are detailed below.

**Simulating fMRI Data for superior deep-learning training: Data augmentation using linear interpolation**

One way to enhance performance of machine learning methods is through use of data augmentation. The basic idea is to use available training sets and generate artificial samples to improve the generalization of a classifier and reduce its variance. In fMRI data analyses, brain regions or voxels can be modeled as vertices in an undirected graph. The correlation among time series of two regions determines the strength of their functional association, which can be represented as the edge weight connecting the two vertices. The functional network obtained from such a graph is considered a complete weighted graph and the set of edge weights is used as the feature vector during classification process. The size of these feature vectors is large, which requires the classifier to optimize a lot of parameters and may cause issues like overfitting. One way to improve the performance of machine-learning and deep-learning techniques is to feed more data to the model to reduce overfitting and improve generalizability. However, MRI acquisition is time consuming and costly, and does not allow strict control of parameters needed for machine-learning algorithmic development.

Therefore, we propose a data-augmentation method which allows us to produce enormous amount of data suitable for training our deep-learning models. Data augmentation techniques have been used for generating synthetic data using available training set for other fields [27] [28] [29] [30] [31]. Our fMRI data augmentation technique is inspired by a technique called SMOTE [32]. SMOTE or Synthetic Minority Over-sampling Technique is an effective model which is used for oversampling the data in minority class in imbalanced datasets. SMOTE generates synthetic data in feature space by using the nearest neighbors of a sample. After k-nearest neighbors of sample A are computed, one of them is selected randomly (A') and synthetic feature vector is computed using the following equation:

$$A' = A + (A - A') \times r$$

In this equation r is a random number which can be different for each feature. In our model, feature vectors will consist of correlation coefficients among different regions. One idea for computing nearest neighbors of a sample could be computing Euclidean distance between correlation arrays but this may not be able to compute the similarity between the complex patterns of fMRI data. In order to compute the similarity between samples and finding nearest neighbors we used a measure called Extended Frobenious Norm (EROS). This measure computes the similarity between two multivariate time series (MTS) [33].

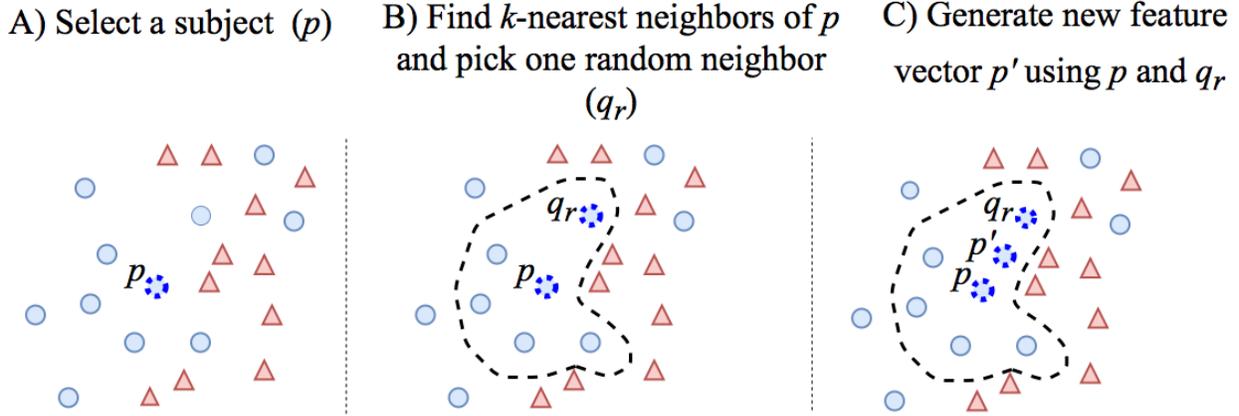

**Figure 1: Selecting artificial data**

The fMRI data consists of several regions each having a time series so we can consider it as a multivariate time series. Our previous study on ADHD disorder has shown that EROS is an effective similarity measure for fMRI data and using it along with k-Nearest-Neighbor achieves high classification accuracy [34]. Eros computes the similarities between two MTS items *A* and *B* based on Eigenvalues and Eigenvectors of their covariance matrix using the following equation.

$$Eros(A, B, w) = \sum_{i=1}^{n} w_i|\langle a_i, b_i\rangle| = \sum_{i=1}^{n} w_i|\cos \theta_i|$$

In this equation $\theta_i$ is the cosine of the angle between ith corresponding Eigenvectors of covariance matrices of A and B. w is the weight vector which is computed based on eigenvalues of all MTS items [33] . The process of computing w includes normalizing eigenvalues of each MTS item followed by normalizing all ith eigenvalues over the entire dataset and normalizing the whole result. Considering EROS as similarity measure and based on the idea of SMOTE, the proposed data augmentation process creates one synthetic sample for each sample in training set which results in doubling the size of the training set. A high-level overview of the augmentation technique is shown in Fig. 1.

Another method that might be useful for the augmentation is by using graph network motifs. Pairwise correlations as features are dependent on the discriminative information that may be available for a given training data set. Some training datasets may contain more noise than others which may lead to sub-optimal data augmentation. Another approach that may be used in conjunction with the approach above is using network motifs which are known to be less prone to random noise. Network motifs reflect functional properties and have wide range of applications. Network motifs are principle subgraphs with particular pattern that are repeated inside a complex network much more than random networks and contain profound information about the network connectivity. One starting point for augmentation in this way is to decompose the brain functional network into its significant building blocks known as Network Motifs in order to unveil the underlying connectivity structures. To do so, we can use different threshold values to remove some connections from the complete functional brain network and on each network compute the frequency of different motifs. Since the edge weights or correlations among regions play an important role in brain disorder diagnosis, besides using motif frequencies which does not consider the edge weights, we incorporate motif intensities and generate feature vectors by concatenating the intensity and frequency of different motifs for each functional network. Next, the data augmentation technique is utilized in order to increase the number of samples. Here the data augmentation is executed by finding the neighboring samples using EROS similarity measure and generating feature vectors by performing linear-interpolation between those motif feature vectors. Using properties of brain functional network motifs as features along with generating synthetic features using proposed data augmentation can provide a robust resource for ML methods.

**Evaluation of the effectiveness of the Augmentation Strategy**

Data augmentation method can be evaluated by monitoring the performance of a deep learning method in presence of new feature vectors and compare it with the performance of the model trained without using augmented features. Statistical criteria such as accuracy, sensitivity, and specificity can be used to compare the performance of a classifier with and without using data augmentation. Increase in accuracy would justify the augmentation of the data and will depict that useful features are being added to the model for better classification.

We expect to develop and test novel data augmentation techniques that will use available structural and functional ASD data to generate artificial samples to improve the generalizability of proposed deep learning solutions. This will yield an analytic approach for determining power and minimum sample size within a machine learning framework.

Once we have methodology to inflate the available MRI data, we can focus on training our machine-learning models that could be used for ASD classification on a spectrum. Note that these ML methods are still in their infancy and are proposed to be used in conjunction with the current

ASD classification methods. However, the long-term goals of these ML models are to be able to classify ASD spectrum in a diagnostic setting without incorporation of other data.

**Existing State of the art Methods:** Classifying subjects with ASD using machine-learning techniques is an active area of research. Iidaka in [35] used probabilistic neural network applied for classifying resting state fMRI (rs-fMRI) data from 312 ASD and 328 healthy control subjects (subjects under 20 years old were selected), which achieved around 90% accuracy. In another work, Plit et al [36] used 178 age- and IQ-matched individuals (89 ASD; 89 TD) and achieved 76.67% accuracy. Chen et al [37] obtained 79.17% accuracy by using support vector machine technique applied on 112 ASD and 128 healthy controls subjects. They investigated the effect of different frequency bands for constructing brain functional network. However, SVM for multiclass applications can be computationally very intensive. Abraham et al [38] achieved 67% classification accuracy on total 871 subjects. For the same set of subjects, Parisot et al [39] proposed a framework based on Graph Convolutional Networks that achieved 70.4% accuracy. In their work they represented the population as a graph in which nodes are defined based on imaging features and phenotypic information describe the edge weights. Khosla et al [40] used 774 subjects containing 379 ASD and 395 healthy controls and achieved 73.4% accuracy by using an ensemble strategy which considers different ROI definitions and using convolutional neural network. Brown et al [41] obtained 68.7% classification accuracy on 1013 subjects including 539 healthy controls and 474 with ASD, by proposing an element-wise layer for deep neural networks which incorporates the data-driven structural priors. Sen et al [42] proposed a new algorithm which combines structural and functional features from MRI and fMRI data and got 64.3% accuracy of classifying 1111 total healthy and ASD subjects. In [43], Neilsen et al obtained 60% accuracy of a group of 964 healthy and ASD subjects. Dvornek et al. [44] achieved 73.2 accuracy in classifying 101 subjects (57.4% ASD subjects) using long-term short-memory network with the addition of generative adversarial network. The work proposed by Kazeminejad et al [45] obtained 95% accuracy in classifying Autistic subjects from healthy samples (subjects with age>30). More recently, Heinsfeld et al [46] used a deep-learning based approach and achieved 70% accuracy for classifying 1035 subjects (505 ASD and 530 controls) and claimed their approach improved the state-of-the-art technique before them which had obtained 67% accuracy. In their technique, distinct pairwise Pearson's correlation coefficients were considered as the features. Two stacked denoising auto-encoders were used as the pre-training stage in order to extract lower dimensional data. After training auto-encoders, their weights were applied to a multi-layer perceptron classifier (fine-tuning process) which was used for the final classification. However, their proposed technique is excessively computer-intensive, and the overall training process took over 32 hours. They also performed the classification for each of the 17 sites included in ABIDE dataset separately and an average accuracy of 52% was reported. They justified this result as there were fewer number of subjects available at each individual site.

**Proposed Machine-learning techniques for ASD Classification**

In this section we will propose and showcase some results that we have obtained by using machine-learning and deep-learning models. We show that using a combination of data-augmentation, correlations metrics and deep-learning techniques can allow much more accurate diagnosis than is currently possible using traditional techniques.

**Feature engineering:** Functional connectivity between brain regions is an important concept in fMRI analysis and is shown to contain discriminative patterns for fMRI classification. Among correlation measures, Pearson's correlation is mostly used for approximating the functional connectivity in fMRI data [47] [48] [49]. It shows the linear relationship between time series of two different regions. Based on symmetric property of Pearson's correlation, we only use the upper triangle (above diagonal) part of the correlation matrix as features for our proposed classification technique. The brain data that is used in our studies is parcellated into 200 regions, hence 19,900 distinct pairwise Pearson's correlations will be considered as our features.

**Deep Learning based model for ASD classification**

**In this section we explain our recent proposed methods, ASD-DiagNet and Auto-ASD-Network in details.**

ASD-DiagNET: A hybrid learning technique for diagnosing Autism Spectrum Disorder using fMRI data

Our recent model, called *ASD-DiagNet* [50], proposes a two-stage dimensionality reduction process to reduce the number of features. In the first step, we select half of the correlations comprising 1/4 largest and 1/4 smallest values and eliminated the rest. To do so, we first compute the average of correlations among all subjects in training set and then pick the indices of the largest positive and negative values from averaged correlation array. We then pick the correlations at those indices from each sample as our feature vector. Keeping half of the correlations and eliminating the rest reduces the size of input features by a factor of 2. Next, we use the proposed data augmentation method to increase the number of training samples. In order to further reduce the number of features from actual and augmented data, we use an auto-encoder to extract the lower-dimensional features from strongest and weakest pairwise functional connections. In an autoencoder, the data is first compressed into a lower dimensional latent-space representation by encoder and decoder tries to reconstruct the original input based on that. Training an auto-encoder is an unsupervised process since it does not use the input labels. The lower dimensional data generated during the encoding process contains useful patterns from the original input data with smaller size and can be used as new features for the classification.

For the classification part, we propose to use a single layer perceptron (SLP) network which uses the bottleneck layer of the auto-encoder as its input. Mostly, when auto-encoders are used as feature extractors, the bottleneck layer of the auto-encoder is used as the input of the following classifier after the auto-encoder is completely trained i.e. it has extracted patterns of the data with fewer dimensions which can be used for reconstructing it with the minimum error. In our proposed method, we try to extract those features which are not only good representation of the data with fewer dimension, but also contain discriminating patterns for classification. To do so, we will involve the classification part in training auto-encoder part as well. Based on our approach, right after computing the latent representation of features by a forward pass in auto-encoder, it is used as the input layer of SLP for classification. The training loss of SLP and auto-encoder are added up to a total loss which is used for training both using back-propagation technique. In this hybrid approach, each network helps the other one to optimize two objectives: unsupervised feature extraction and supervised classification. The proposed process is shown in Fig. 2.

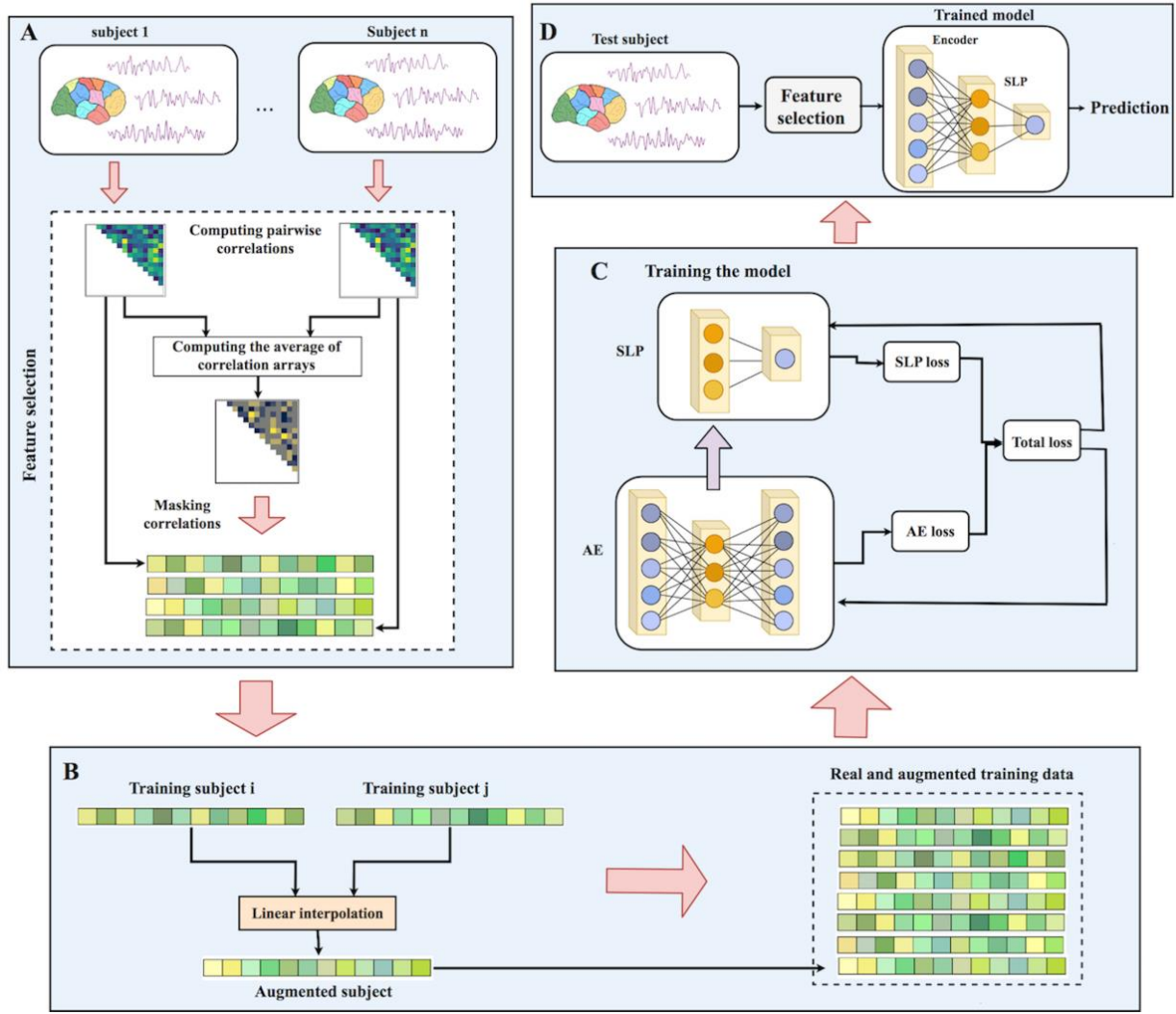

**Figure 2: workflow of ASD-DiagNet**

**Classification and hyperparameter tuning:**

Hidden layers in the deep-learning model learn from the input data and the final layer in the model receives the extracted pattern and features from the previous layers and performs the classifications. We hypothesized that using deep-learning features along with traditional SVM methods could result in higher accuracy than is possible using a single model. In addition to better accuracy, SVM is more explainable and could help us figure out the complex network in ASD brains. SVM classifier has different set of hyperparameters e.g. penalty parameter, kernel function and specific kernels parameters etc. In order to find the optimal hyperparameters for SVM, we used a tool called Auto-Tuned Models (ATM) [51] which allows us to automate and optimize the hyperparameter tuning process. ATM implements the paramerter-search by partitioning the hyperparameter-space using conditional-parameter tree (CPT) where each branch of CPT is a fixed hyper-partition. ATM is configured by using a budget for resource allocation which can either be computation time (max. amount of time spent searching for the parameter) or total number of classifiers to try (which we set to 50 classifiers). We used the top 10 classifiers with highest cross-validation accuracy on the training set and performed an ensemble voting mechanism to predict the label. The workflow of Auto-ASD-Network is shown in Fig. 3.

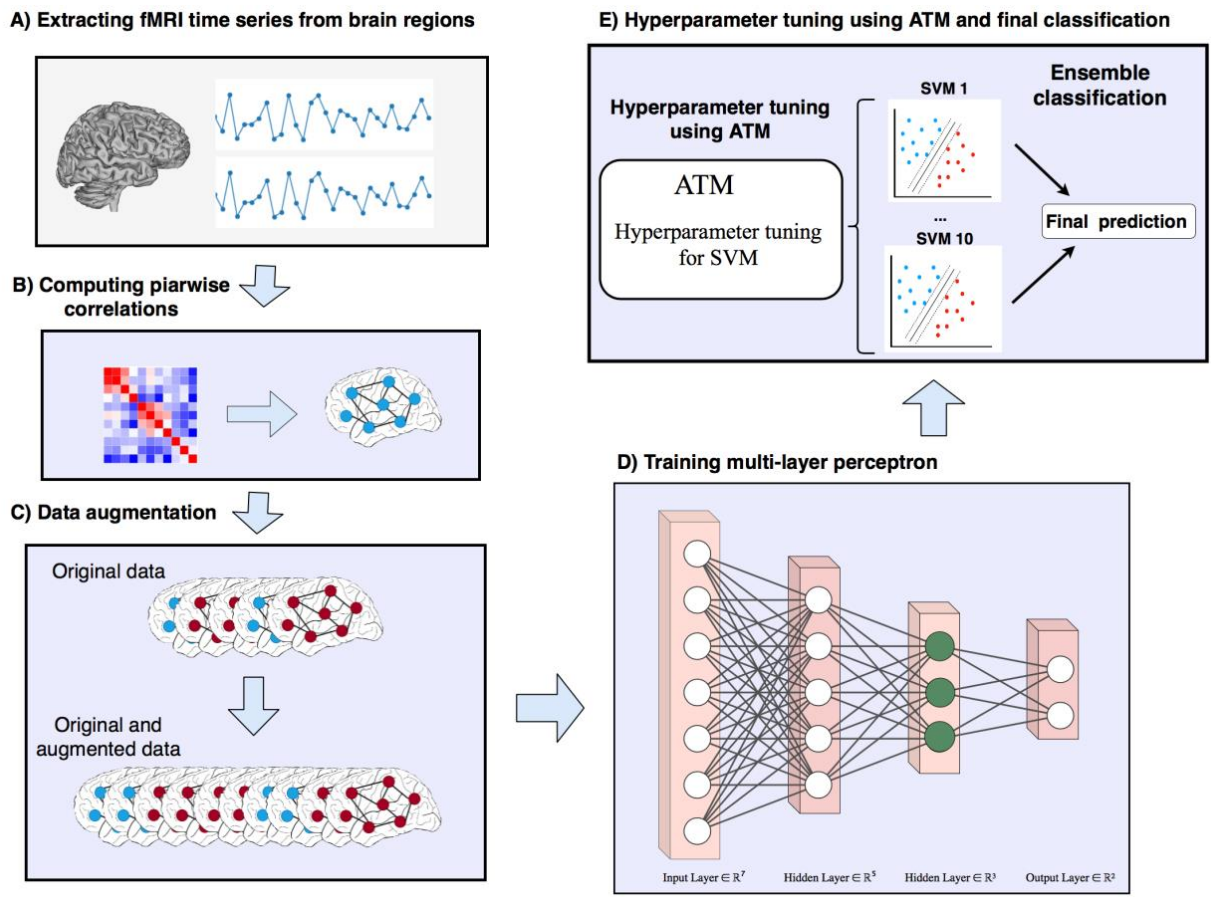

Figure 3: Workflow of Auto-ASD-Network

## Significant Results

**In this section we report the performance of our proposed methods, ASD-DiagNet and Auto-ASD-Network on four public datasets provided by ABIDE initiative. A multi-layer perceptron with two hidden layers and the method proposed in** [46] are used **as the**

baselines to compare with our method. The accuracy, sensitivity and specificity of each approach is reported in table 1.

| Site | Method | Accuracy | Sensitivity | Specificity |
|------|--------|----------|-------------|-------------|
| OHSU | MLP | 64 | 62.5 | 61.6 |
|      | Ref [46] | 74 | 66.6 | 86.6 |
|      | ASD-DiagNet | 82 | 79.1 | 83.3 |
|      | Auto-ASD-Network | 80 | 73 | 83 |
| NYU  | MLP | 68.5 | 44 | 87 |
|      | Ref [46] | 64.5 | 78 | 46 |
|      | ASD-DiagNet | 68 | 50 | 71 |
|      | Auto-ASD-Network | 70 | 57.9 | 79.2 |
| USM  | MLP | 64 | 100 | 0 |
|      | Ref [46] | 62 | 20 | 84 |
|      | ASD-DiagNet | 68.2 | 93.4 | 22 |
|      | Auto-ASD-Net | 72.4 | 87.3 | 45 |
| UCLA | MLP | 71.9 | 76.7 | 64.8 |
|      | Ref [46] | 57.7 | 58 | 57.4 |
|      | ASD-DiagNet | 73.2 | 82.3 | 61.9 |
|      | Auto-ASD-Network | 72.2 | 82.3 | 59.8 |

As can be observed, for all four datasets our proposed methods achieved higher accuracy than other techniques with highest accuracy of 82% on OHSU dataset.

**Future Work: Integration of fMRI and sMRI data for greater classification accuracy}**

Our machine-learning and deep-learning models have been able to get more than 82% accuracy on classifications of ASD and healthy subjects. However, there is still room for improvement to reach the classification accuracy that surpasses the clinical assessments.

**Machine-learning for s-MRI data:** To this end, we will utilize sMRI data and design efficient machine learning methods and deep learning models which will be trained on a set of anatomical features. We will consider various cortical parcellation atlases like Destrieux, Desikan-Killiany, DKT, and automated anatomical labeling (AAL) for parcellating the cortical cortex into regions of interest. Morphometric information such as the number of vertices, surface area, gray matter volume, average thickness, integrated rectified mean curvature and integrated rectified gaussian curvature of the surface from anatomical regions will be extracted as its features. These attributes will be aggregated as the feature vector of each brain. In order to find out which properties contain more discriminating features; recursive feature elimination will be used to select best possible properties and use them for the classification part. The selected values are then used for training classifiers such as SVM, and random forest, etc. Besides using morphological properties as feature vectors, they can be used for generating morphological brain network. In this network each region corresponds to a node and the similarity between morphological properties indicate the strength of the connection between them. Several methods based on Euclidean distance [52], KL-divergance [53] and Pearson's correlation [54] will be investigated. Weights of these graphs can be considered as a feature vector or as an image. We will explore different deep learning methods such as CNN and MLP for classification of subjects using these networks.

**Integration of fMRI and sMRI:** We will design and implement deep learning-based methods that combine the power of functional and structural MRI for diagnosing brain disorders. Providing structural and functional information at the same time, enables the model to extract important discriminative patterns that cannot be detected by using each source of data separately. Most of the proposed models for diagnosing brain disorders are designed based on one source of data and overlook the information provided by other modalities. Our method will work with the brain functional and morphological networks corresponding to the functional and structural data. Both of these networks have shown promising results for diagnosing brain disorders such as ASD, ADHD, and Alzheimer's individually, but their combinatory power has not yet been explored.

Constructing morphological and functional brain connectivity networks: The first step of the algorithm is generating functional and morphological networks. In either of these networks, a node corresponds to one region of interest (ROI) of the brain. The edges refer to functional connectivity and similarity between anatomical features of two regions in functional and morphological network respectively.

**Hybrid fMRI/sMRI learning method:** We will use the weights of the edges of the functional and morphological networks as the feature vector of each modality. Considering n and n' as the number of parcellations in fMRI and sMRI data respectively, functional and morphological networks will contain $\frac{n(n-1)}{2}$ and $\frac{n'(n'-1)}{2}$ distinct pairwise connections. These numbers are usually larger than available number of samples which can cause issues like curse of dimensionality and overfitting. In order to reduce the number of features, we will explore feature extraction methods such as Autoencoders. We will use two separate Autoencoders for extracting low dimensional and high-level feature vectors from functional and morphological connections. After the training process of each autoencoder is ended, we will extract the features from the bottleneck of each autoencoder and append them to generate a *joint feature vector*. At this point, we add the phenotypic information of each subject to the fused feature vector to provide more knowledge about each subject to the classifier. Finally, we will train a multi-layer perceptron as the final classifier. The last layer this network will contain two nodes each providing the probability of test sample belonging to class of healthy or patient.

## **Conclusions**

Our research brings applications from multiple disciplines such as computational sciences, neuroscience, psychiatry, machine-learning algorithms, and high-performance computing. Innovative deep-learning algorithms are proposed and HPC solutions will have significant improvement in accuracy and time of big fMRI data analysis. Our interdisciplinary approach will help in quantifying the current psychiatric diagnosis and is based on machine-learning and deep learning algorithms which can increase the accuracy of diagnosis, prognosis, and treatments of difficult to assess mental disorders such as ADHD and ASD. Our machine-learning algorithms can be helpful in identifying mental disorders in children where the symptoms of mental health disorder may have a delayed onset or cases where the ASD symptomology might not be very sharp. Such early detection in ASD and subsequent intervention can significantly improve a child's mental health. Further, such classification can be instrumental in identifying mental disorders for incarcerated population which may otherwise be prone to implicit biases by mental health practitioners. Therefore, our techniques will likely have a significant difference in identifying mental disorders for vulnerable population segments. Our novel machine-learning models will also help reveal previously unobservable phenomenon of mental-disorder features at a much finer scale than is currently possible. Rapid analysis of big data using ubiquitous architectures with improved efficiency and reduced costs will move us one step closer to personalized medicine. Our proposed machine-learning algorithms are akin to blood-tests that were developed in the 19$_{th}$ century to quantify the diagnosis and prognosis of many diseases. We are confident that our algorithms will be a significant milestone for diagnosing and identifying mental disorders using imaging and explainable machine-learning models.